\documentclass[runningheads,natbib]{svjour2}
\bibpunct{[}{]}{;}{n}{}{,} 
\smartqed  
\usepackage{amsmath,amsfonts,amssymb,mathrsfs,graphicx}

\newcommand{\bi}{\begin{itemize}}
\newcommand{\ei}{\end{itemize}}

\newcommand{\be}{\begin{equation}}
\newcommand{\ee}{\end{equation}}
\newcommand{\bea}{\begin{eqnarray}}
\newcommand{\eea}{\end{eqnarray}}

\newcommand{\deltacp}{\delta_{\mathrm{CP}}}
\newcommand{\stheta}{\sin^2(2 \theta_{13})}

\newcommand{\ie}{{\it i.e.}}

\newcommand{\eg}{{\it e.g.}}

\newcommand{\cf}{{\it cf.}}

\newcommand{\eq}{Eq.}

\newcommand{\fig}{Fig.}

\newcommand{\Ref}{Ref.}
\newcommand{\Refs}{Refs.}

\newcommand{\equ}[1]{\eq~(\ref{equ:#1})}
\newcommand{\figu}[1]{\fig~\ref{fig:#1}}

\journalname{Earth, Moon, and Planets}
\begin{document}

\title{Neutrino tomography \thanks{Supported by the W.~M.~Keck Foundation and NSF grant PHY-0503584.}
}
\subtitle{Learning about the Earth's interior using the propagation of neutrinos}


\author{Walter Winter}


\institute{Walter Winter \at
              Institute for Advanced Study \\
	      School of Natural Sciences \\
	      Einstein Drive \\
	      Princeton, NJ 08540 \\
              \email{winter@ias.edu}           
}

\date{Received: }

\maketitle

\begin{abstract}
Because the propagation of neutrinos is affected by the presence of Earth matter, it opens new possibilities to probe the Earth's interior. Different approaches range from techniques based upon the interaction of high energy (above TeV) neutrinos with Earth matter, to methods using the MSW effect on the oscillations of low energy (MeV to GeV) neutrinos. In principle, neutrinos from many different sources (sun, atmosphere, supernovae, beams etc.) can be used. In this talk, we summarize and compare different approaches with an emphasis on more recent developments.
In addition, we point out other geophysical aspects relevant for neutrino oscillations.

\end{abstract}

\newpage

\section{Introduction}

Neutrinos are elementary particles coming in three
active (\ie, weakly interacting) flavors. Since the
cross sections for neutrino interactions are very small,
neutrinos practically penetrate everything. However, one can compensate
for these tiny cross sections by just using enough material
in the detector. Depending on neutrino energy and source, the 
detector has to be protected from backgrounds such that
the neutrino events cannot be easily mixed up with different particle 
interactions. Neutrinos are produced 
in detectable numbers and with detectable energies 
by nuclear reactions in the sun, by cosmic
ray interactions in the Earth's atmosphere, in nuclear fission reactors,
in supernova explosions, in the Earth's crust, 
and possibly by astrophysical sources. In addition, accelerator-based
neutrino sources  specifically designed to produce a high-intensity 
neutrino beam have been successfully operated (such as K2K~\cite{Aliu:2004sq} or MINOS)
 or are planned.
Thus, there are neutrinos from various different sources with different energies.

One of the most recent exciting discoveries in neutrino physics
is neutrino oscillations, \ie, neutrinos change flavor while
traveling from source to detector. This quantum mechanical phenomenon
implies that neutrinos mix, \ie, the eigenstates of the weak interaction
are not the same as the mass eigenstates, and at least two out of the three
have non-vanishing masses. This is probably the most direct evidence today for
physics beyond the standard model of elementary particle physics.
Recent neutrino oscillation experiments, especially SNO~\cite{Ahmad:2002jz}, KamLAND~\cite{Eguchi:2002dm},
Super-Kamiokande~\cite{Fukuda:1998mi}, and CHOOZ~\cite{Apollonio:1999ae} have helped 
to quantify this picture. Unlike in quark 
mixing, two out of the three mixing angles are large, and one is even close
to maximal. In addition, the oscillation 
frequencies have been fairly precisely measured.
For one of the mixing angles $\theta_{13}$, however, only an upper bound exists, and
several parameters (the arrangement of masses, \ie, mass hierarchy, and one complex
phase $\delta_{\mathrm{CP}}$ relevant for neutrino oscillations) are still unknown. 
Future experiments will probe these parameters starting with the Double Chooz~\cite{Ardellier:2004ui},
T2K~\cite{Itow:2001ee}, and NO$\nu$A~\cite{Ayres:2004js} experiments (for the prospects for the next decade, see, \eg, \Ref~\cite{Huber:2004ug}).

For neutrino tomography the relevant aspect is the sensitivity to Earth
matter. Since it is well known that the cross sections with matter rise
at least until $10 \, \mathrm{TeV}$~\cite{Quigg:1986mb}, the probability of matter interactions
can be increased by higher neutrino energies. Neutrino absorption tomography
uses this effect to infer on the matter structure. For neutrino oscillations,
we know that the so-called MSW matter effect~\cite{Wolfenstein:1978ue,Mikheev:1985gs,Mikheev:1986wj} is the most plausible explanation for
the solar neutrino deficit~\cite{Fogli:2004zn}. This, however, implies that neutrino oscillations
in the Earth have to experience this effect, too. Neutrino oscillation tomography
uses the MSW effect to study the matter structure.

\section{Tomography using the propagation of neutrinos}

Tomography using the propagation of neutrinos~\cite{Nedyalkov:1981yy,Nedyalkov:2}
assumes a neutrino source
with a well-known flux and flavor composition, a well-understood
neutrino detector, and a specific neutrino propagation model
between source and detector. The key ingredient to any such
tomography is a considerable dependence of the propagation model 
on the matter structure between source and detector. 
Compared to the detection
of geoneutrinos, the object of interest is not the neutrino
source, but the material along the baseline (path between source 
and detector). If the matter structure along the baseline
is (partly) unknown, the information from counting neutrino 
events at different energies by the detector can 
be used to infer on the matter profile.
Two accepted propagation models could be used for
neutrino tomography:
\begin{description}
\item[Neutrino absorption:] Because the cross section for
neutrino interactions increases proportional to the energy,
neutrino interactions lead to attenuation effects. Useful
neutrino energies for a significant attenuation are $E_{\nu} \gtrsim 1 \, \mathrm{TeV}$.
\item[Neutrino oscillations:] The MSW effect~\cite{Wolfenstein:1978ue,Mikheev:1985gs,Mikheev:1986wj}
in neutrino oscillations (coherent forward scattering in matter)
leads to a relative phase shift of the electron flavor
compared to the muon and tau flavors. This phase shift depends on the electron density.
Useful neutrino energies require substantial contributions from the MSW effect
as well as large enough oscillation amplitudes. Depending the relevant $\Delta m^2$, neutrino
energies between $100 \, \mathrm{MeV}$ and $35 \, \mathrm{GeV}$ are optimal for
studying the Earth's interior.
\end{description}
Beyond these two models, at least small ad-mixtures of non-standard effects have not
yet been excluded. Some of these non-standard effects are sensitive to the matter
density, too. Examples are mass-varying neutrinos with acceleron couplings to matter fields~\cite{Kaplan:2004dq}, non-standard neutrino interactions (see \Ref~\cite{Huber:2002bi}
and references therein),
and matter-induced (fast) neutrino decay~\cite{Giunti:1992sy}. Because there
is not yet any evidence for such effects, we do not include them in this discussion.

Given the above neutrino energies, there are a number of potential sources which could
be used for neutrino propagation tomography. For neutrino oscillations, solar neutrinos,
supernova neutrinos, atmospheric neutrinos, and neutrino beams (such as superbeams or
neutrino factories) are potential sources. For neutrino absorption, high-energy 
atmospheric neutrinos, a possible high-energy neutrino beam, or cosmic sources
are possible sources.

As far as potential geophysics applications are concerned, neutrinos may be interesting
for several reasons:
\begin{enumerate}
\item 
 Neutrinos propagate on straight lines. The uncertainty in their path (direction) 
 is only as big as the surface area of the detector.
\item 
 Neutrinos are sensitive to complementary quantities to geophysics: Neutrino absorption
 is directly sensitive to the matter density via the nucleon density. Neutrino oscillations
 are sensitive to the electron density which can be converted in the matter density
 by the number of electrons per nucleon (for stable ``heavy'' materials about two).
 On the other hand, seismic wave geophysics needs to reconstruct the matter density by the equation
 of state from the propagation velocity profile.
\item 
 Neutrinos are, in principle, sensitive to the density averaged over the baseline, whereas
 other geophysics techniques are, in principle, less sensitive towards the innermost 
 parts of the Earth. For example, seismic shear waves cannot propagate within the outer
 liquid core, which means that a substantial fraction of the energy deposited
 in seismic waves is reflected at the mantle-core boundary. Other direct density
 measurements by the Earth's mass or rotational inertia are less sensitive towards
 the innermost parts, too, because they measure volume-averaged quantities.
\end{enumerate}
Given these observations, there may be interesting geophysics applications exactly where
complementary information is needed. Possible applications range from the
detection of density contrasts in the Earth's upper or lower mantle, to 
the measurement of the average densities of the outer and inner core by
independent methods.

\section{Neutrino absorption tomography}

Here we discuss tomography based on attenuation effects
in a neutrino flux of high enough energies, which we call,
for simplicity, ``neutrino absorption tomography''. After we
have introduced the principles, we will discuss possible
applications with respect to tomography of the whole
Earth as well as specific sites.

\subsection{Principles}

``Neutrino absorption tomography'' uses the attenuation of
a high-energy neutrino flux as a propagation model.
In this case, weak interactions damp the initial flux
by the integrated effect of absorption, deflection, 
and regeneration. For example, muons produced by a muon
neutrino interaction are absorbed very quickly in Earth
matter, whereas tauons produced by tau neutrinos tend to decay before
absorption (and some of the decay products are again neutrinos). 
Only the integrated effect leads to attenuation
of the flux. The magnitude of the attenuation effect can be estimated from
the cross section
\begin{equation}
\frac{\sigma}{E} \sim 10^{-35} \, \frac{\mathrm{cm^2}}{\mathrm{TeV}}
\label{equ:crossx}
\end{equation}
to be of the order of several per cent over the Earth's diameter 
for $E_\nu=1 \, \mathrm{TeV}$.
The interaction cross section rises linearly up to about 
$10 \, \mathrm{TeV}$~\cite{Quigg:1986mb}, whereas the behavior
above these energies is somewhat more speculative.
The energies are usually as high as standard 
neutrino oscillations  do not develop within the Earth. 
Since the neutrinos interact with nucleons, the attenuation
is directly proportional to the nucleon density. Therefore,
neutrino absorption is a very directly handle on the matter
density with an extremely tiny remaining uncertainty from
composition and the difference between neutron and proton mass.

As far as possible potential neutrino sources are concerned,
\equ{crossx} requires very high neutrino energies. The
existence of corresponding neutrino sources is plausible and will
be tested by upcoming experiments commonly referred to as 
``neutrino telescopes''. These neutrino telescopes could
also serve as prototypes for the detectors useful for
neutrino absorption tomography. The only detected source so far
is atmospheric neutrinos produced by the interaction of
cosmic rays in the atmosphere. Unfortunately, the atmospheric
neutrino flux drops rapidly with energy, which means that
statistics is limited at the relevant energies $E_\nu > 1 \, \mathrm{TeV}$ 
(see, \eg, \Ref~\cite{Gonzalez-Garcia:2005xw} for specific
values). Other potential candidates include many different possible
astrophysics objects, as well as particle physics mechanisms 
such as decays of dark matter particles. Since we know that the 
Earth is hit by cosmic rays of very high energies, it might be inferred that
astrophysical mechanisms exist which accelerate particles 
(for example, protons) to these high energies. It is 
plausible that such mechanisms also produce neutrinos. 
Potential mechanisms could either produce
discrete fluxes from individual objects, or their integrated
effect could lead to a diffuse flux over the whole sky.
Eventually, one could think about a neutrino beam producing
high-energy neutrinos. If, for instance, one used the protons
from LHC ($7 \, \mathrm{TeV}$) to hit a target, the decaying 
secondaries (pions, kaons) would produce a neutrino flux peaking 
at about $1 \, \mathrm{TeV}$.

\subsection{Whole Earth tomography}

\begin{figure}[t]
\begin{center}
\begin{tabular}{ccc}
\includegraphics[width=3cm]{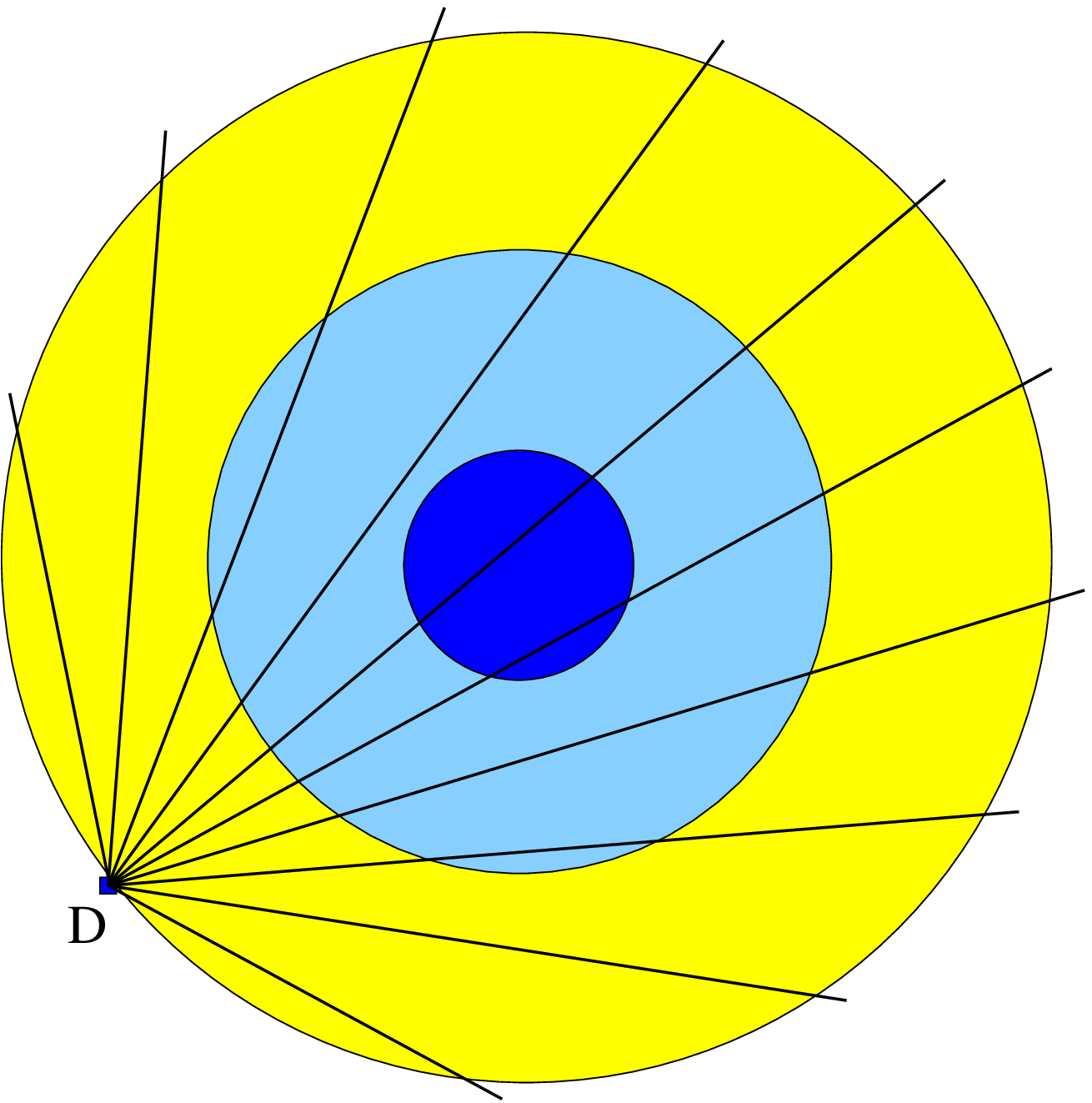} & 
\includegraphics[width=3cm]{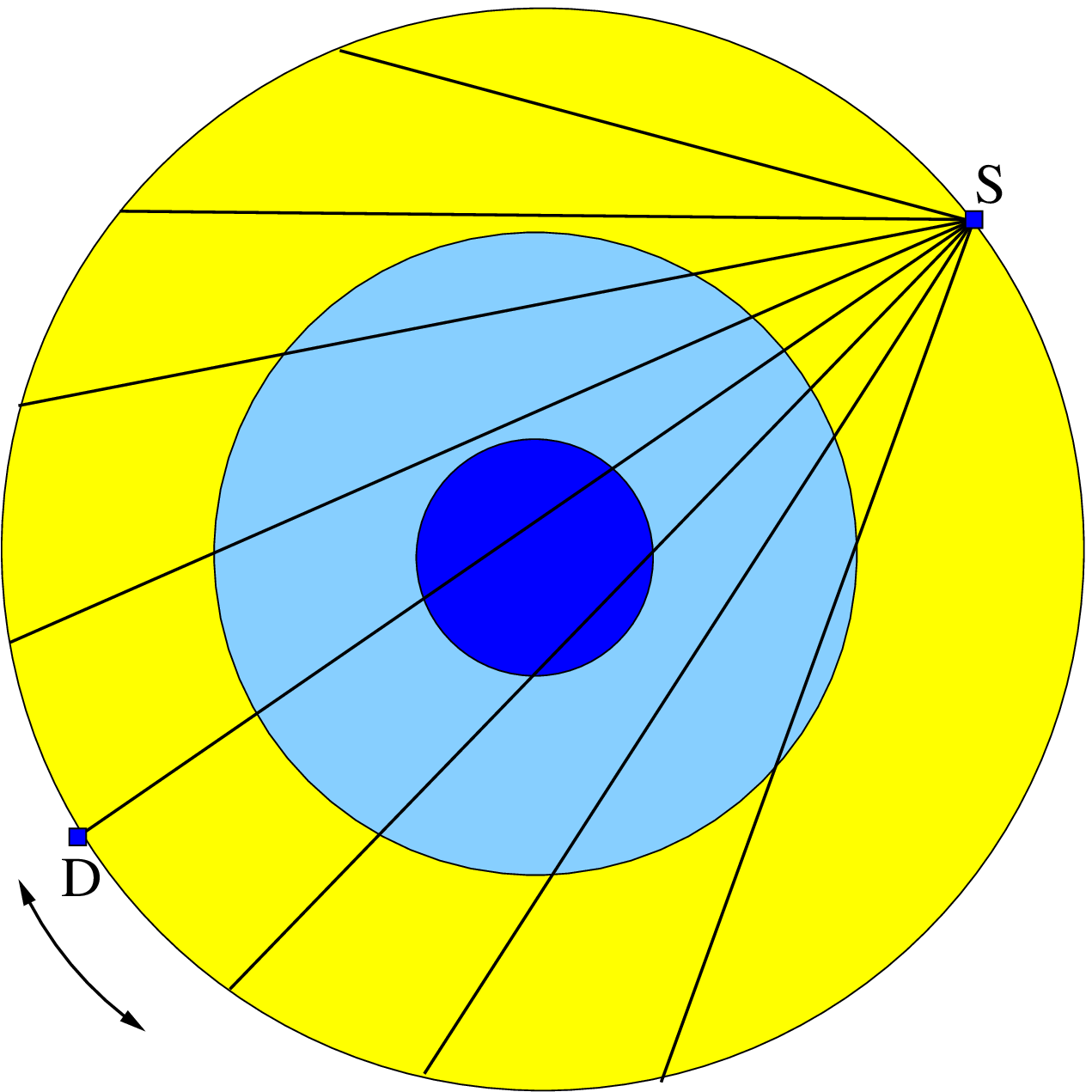} & 
\includegraphics[width=3.6cm]{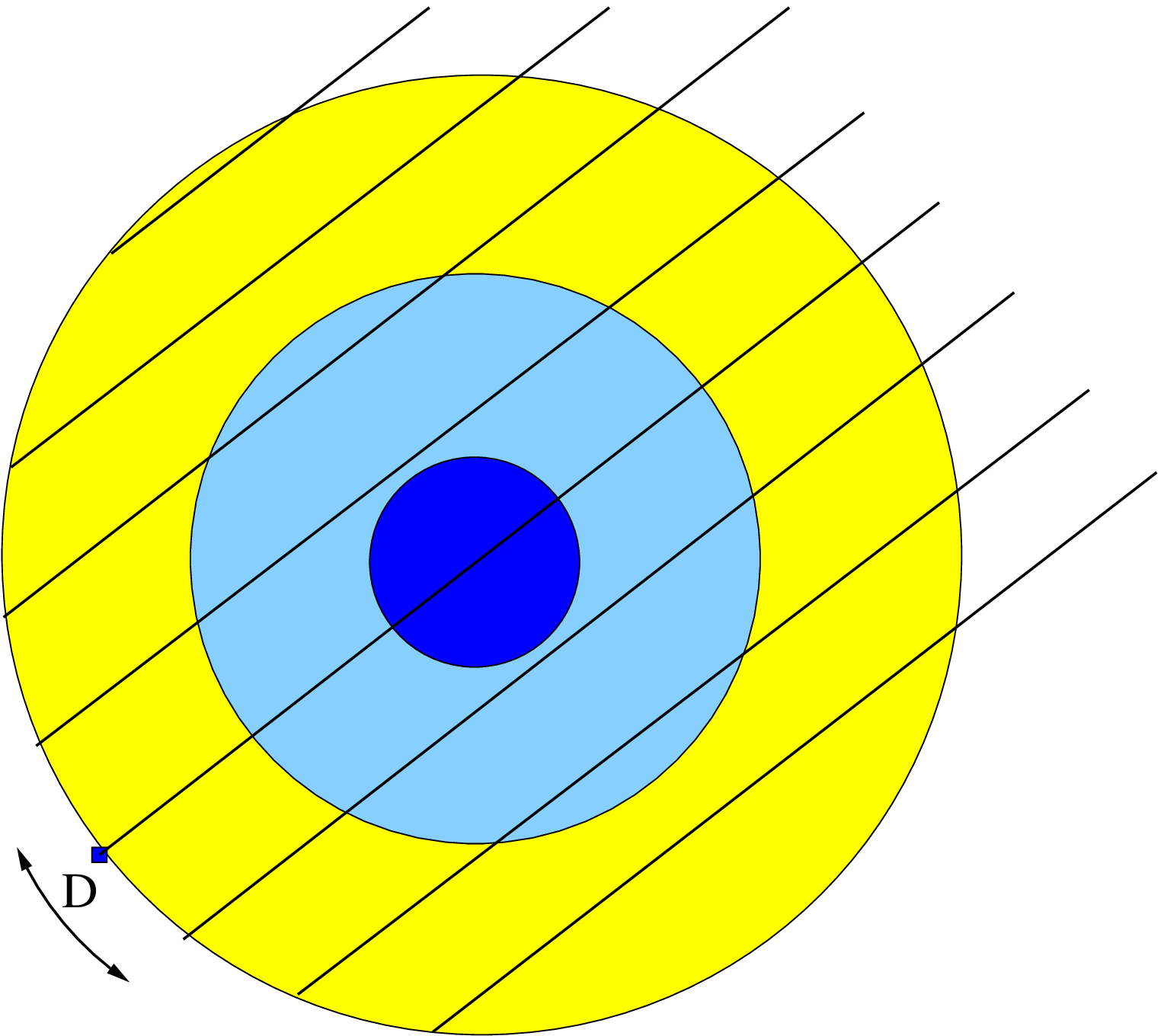} \\[0.3cm]
a) Isotropic flux &  b) High-energy  &
 c) Cosmic point  \\
 & neutrino beam & source
\end{tabular}
\end{center}
\caption{\label{fig:wholeearth} Three different approaches
to ``Whole Earth Tomography'' using neutrino absorption. The
lines refer to different baselines.
}
\end{figure}

For possible applications of neutrino absorption tomography, 
there exist two different directions in the literature:
Either one could ``X-ray'' the whole Earth (``Whole Earth tomography''),
or one could think about the investigation of specific sites in the
Earth's mantle. We summarize in \figu{wholeearth} different approaches
to ``Whole Earth tomography''. In case a) (isotropic flux) a neutrino
flux from many directions is detected by a detector with good directional
resolution. For instance, a possible neutrino source would be a
cosmic diffuse flux~\cite{Jain:1999kp} (related work: \Ref~\cite{Reynoso:2004dt}) 
or the high-energy tail of
atmospheric neutrinos (see, \eg , \Ref~\cite{Gonzalez-Garcia:2005xw}).
This application could be very interesting because it might be available
at no additional experimental effort. However, if one wants to study the
innermost parts of the Earth, it is (except from sufficient directional
resolution and flux isotropy) a major challenge that the fraction of the sky which is seen through 
the Earth's inner core is very small
($\sim1\%$), which means that the statistics for this specific goal is very low.
Very good precisions may, on the other hand, be obtained for the mantle (see \fig~4
of \Ref~\cite{Jain:1999kp}). In case b) (high-energy neutrino beam)~\cite{DeRujula:1983ya,Wilson:1983an,Askar84,Borisov:1986sm,Borisov:1989kh} the
detector is moved to obtain many baselines, whereas the source is kept fixed.
In this case, high precisions could be obtained~\cite{DeRujula:1983ya}. However, a major challenge
might actually be the operation of a high-energy neutrino beam
with a moving decay tunnel. Note that such a beam could not only be used
for whole Earth tomography, but also for local searches (see below).
In case c) (cosmic point source)~\cite{Wilson:1983an,Kuo95}, the flux from a single object is used
for the tomography of the Earth. In this case, the flux has to be constant in time
to be detected either by a moving detector, or by one detector using many baselines
by the rotation of the Earth. Note that the second mechanism cannot be used for
the currently largest planned neutrino telescope ``IceCube''~\cite{Ahrens:2003ix} 
because it is residing at the south pole.

\subsection{Specific site tomography}

\begin{figure}[t]
\begin{center}
\includegraphics[width=\textwidth]{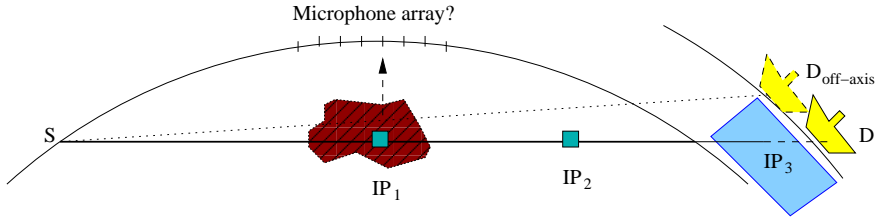}
\end{center}
\caption{\label{fig:tevbeam} Different possibilities for
neutrino tomography using a high-energy neutrino beam.
The labels ``IP'' refer to possible interaction points.
See text for more details.
}
\end{figure}

Compared to ``whole Earth tomography'', a different direction is the investigation of
individual sites, such as in the Earth's mantle. For example, \Ref~\cite{DeRujula:1983ya}
extensively reviews techniques based on a high-energy neutrino beam. We summarize
some of those in \figu{tevbeam}. The neutrinos, produced by the source ``S'', may interact
at several possible interaction points $IP$. If, for example, the site of
interest is the dark-shaded cavity, an interaction at $IP_1$ could create a particle shower
leading to sound production, which may be detected by a microphone array at the surface.
In addition, the final neutrino flux detected at ``D'' would be damped depending on
the material density in the cavity. An interaction at $IP_2$ just below the surface ($\lesssim 200 \, \mathrm{m}$) would produce muons which could still be detected at the surface (such as 
possibly by a muon detector on a truck). A variation of this flux detected
by a moving muon detector could point towards heavy materials. Eventually, a neutrino 
interaction at $IP_3$ within the sea water below a muon moving detector would indicate that
the initial neutrino has arrived. Since the neutrino energy decreases rapidly by moving the
detector out of the beam axis by kinematics, attenuation effects also decrease and the initial flux could be measured by the ``off-axis'' technology. Comparing this flux to the on-axis flux reveals the attenuation along the path and therefore some information on the matter structure. 

In summary, there are many potential applications of neutrino absorption tomography. The coming
years, especialy the operation of IceCube, will reveal the possible existence of high-energy
cosmic neutrino fluxes. Operating a high-energy neutrino beam may be a major technical
challenge, which definitively needs further investigation.

\section{Neutrino oscillation tomography}

In this section, we discuss neutrino tomography using oscillations. First, we
introduce the principles of neutrino oscillation tomography: Neutrino oscillations
in vacuum and matter, numerical approaches to neutrino oscillation tomography,
as well as conceptual (mathematical) problems. Then, we show
applications related to solar and supernova neutrinos, and we 
discuss tomography with neutrino beams.

\subsection{Principles}

Neutrino oscillation tomography uses neutrino oscillations in matter as
propagation model. Possible neutrino sources include ``natural'' ones (\eg, sun, supernova,
atmosphere), as well as ``man-made'' ones (\eg, superbeam, $\beta$-beam, neutrino factory).
The detection technology depends on the neutrino energy and ranges from water Cherenkov
detectors (lower energies), over liquid scintillators (medium energies), to iron
calorimeters (high energies), just to mention some examples. 

\subsubsection*{Neutrino oscillation phenomenon}

Neutrino oscillations
are a quantum mechanical phenomenon with two prerequisites: First, the weak interaction eigenstates
have to be different from the propagation/mass eigenstates (flavor mixing). Second, the
neutrino masses have to be different from each other, which implies that at least two of the 
active neutrinos
have to have non-zero mass~\cite{Bilenky:1978nj}. In the limit of two flavors, the flavor transition probability $\nu_\alpha \rightarrow \nu_\beta$ in vacuum can be written as
\begin{equation}
P_{\alpha \beta} = \sin^2 ( 2 \theta ) \, \sin^2 \left( \frac{\Delta m^2 L}{4 E} \right) \, ,
\label{equ:oscvac}
\end{equation}
where $\theta$ is the mixing angle of a $2 \times 2$ rotation matrix $U$, $\Delta m^2 \equiv m_a^2 - m_b^2$ is the mass-squared difference describing the oscillation frequency, $L$ is the
baseline (distance source-detector), and $E$ is the neutrino energy. Note that the
quotient $L/E$ determines the oscillation phase. Similarly, the flavor conservation probability $\nu_\alpha \rightarrow \nu_\alpha$ is given by $P_{\alpha \alpha} = 1-P_{\alpha \beta}$
from conservation of unitarity. Practically, $P_{\alpha \beta}$ is measured
as function of $E$ (convoluted with the neutrino flux and cross sections)
 for a fixed baseline since the detector 
cannot be moved. Since we do know that we deal with three active flavors, the complete picture
is somewhat more complicated. Three-flavor neutrino oscillations can be described by six parameters (three mixing angles, one complex phase, and two mass squared differences), which decouple
into two-flavor oscillations, described by two parameters each, in certain limits
(see \Ref~\cite{Fogli:2005cq} for a recent review). In summary, we have two almost decoupled
two-flavor oscillations described by two very different frequencies and large mixing angles,
often referred two as ``solar'' ($\Delta m_{21}^2$, $\theta_{12}$) and ``atmospheric''
($\Delta m_{31}^2$, $\theta_{23}$) oscillations. Those could be coupled by $\theta_{13}$, for
which so far only an upper bound $\sin^2 (2 \theta_{13}) \lesssim 0.1$~\cite{Apollonio:1999ae} exists. In addition,
we do not yet know anything about the complex phase $\deltacp$, which could lead to sub-leading effects, and the sign of $\Delta m_{31}^2$ (``mass hierarchy''). These parameters will be probed by neutrino oscillation experiments in the coming years. In this section, we concentrate on the two-flavor case for pedagogical reasons.

\subsubsection*{Matter effects in neutrino oscillations}

Key ingredient to neutrino tomography are matter effects in neutrino oscillations~\cite{Wolfenstein:1978ue,Mikheev:1985gs,Mikheev:1986wj}. Since Earth matter
contains plenty of electrons, but no muons or tauons, charged-current interactions
of the electron neutrino flavor through coherent forward scattering lead to a relative 
phase shift compared to the muon and tau neutrino flavors. In the Hamiltonian in two flavors,
the matter term enters as the second term in
\begin{equation}
\mathcal{H}(n_e) = U \left( \begin{array}{cc}
0 & 0 \\
0 & \frac{\Delta m_{21}^2}{2E} 
\end{array}
 \right) U^\dagger + \left( \begin{array}{cc}
A(n_e) & 0 \\
0 & 0
\end{array}
 \right)
 \label{equ:h}
\end{equation}
in flavor space, where $A(n_e) = \pm \sqrt{2} G_F n_e$ is the matter potential
as function of the electron density $n_e$ and the coupling constant $G_F$, and the different signs
refer to neutrinos (plus) and antineutrinos (minus). Assuming that the number of electrons per nucleon is approximately $0.5$ for stable ``heavy'' (considerably heavier than hydrogen) materials,
the electron density can be converted into the matter density as $n_e = 0.5 \,  \rho/m_N$ with $m_N$ the nucleon mass. In this case, there is some material dependence of this factor $0.5$ (``electron fraction''), which, however, might also be used to obtain additional information on the composition.
In two flavors and for constant matter density, \equ{oscvac} can be easily re-written
by a parameter mapping between vacuum and matter parameters:
\begin{equation}
P_{\alpha \beta} = \sin^2 ( 2 \tilde{\theta} ) \, \sin^2 \left( \frac{\Delta \tilde{m}^2 L}{4 E} \right)\, ,
\label{equ:oscmatter}
\end{equation}
where
\begin{equation}
\Delta \tilde{m}^2 =  \xi \cdot \Delta m^2 \, ,  \quad 
\sin (2 \tilde{\theta})  =  \frac{\sin (2 \tilde{\theta})}{\xi} \, ,
\label{equ:mapping}
\end{equation}
with
\begin{eqnarray}
\xi & \equiv & \sqrt{\sin^2 (2 \theta) + (\cos (2 \theta) - \hat{A})^2} \, , 
\label{equ:xi} \\
\hat{A} & \equiv &  \frac{2 \sqrt{2} G_F n_e E}{\Delta m^2} \, .
\label{equ:ahat}
\end{eqnarray}
One can easily read off these formulas that for $\hat{A} \rightarrow \cos (2 \theta)$
the parameter $\xi$ in \equ{xi} becomes minimal, which means that the oscillation frequency
in matter becomes minimal and the effective mixing maximal (\cf, \equ{mapping}).
This case is often referred to as ``matter resonance'', where the condition $\hat{A} \rightarrow \cos (2 \theta)$ evaluates to 
\begin{equation}
E_{\mathrm{res}}  \sim 13 \, 200 \, \cos(2 \theta) \, \frac{\Delta m^2 \, [\mathrm{eV^2}]}{\rho \, [\mathrm{g/cm^3}]} \, .
\label{equ:eres}
\end{equation}
This condition together with the requirement of a large oscillation phase $\sin^2 ( \Delta m^2 L/(4 E) ) = \mathcal{O}(1)$ leads to the ``ideal'' energies for neutrino oscillation tomography
depending on the considered $\Delta m^2$:
\begin{eqnarray}
\Delta m_{21}^2: & & E \sim 100 \, \mathrm{MeV} \, \mathrm{to} \, 1 \, \mathrm{GeV} \, , \nonumber \\
\Delta m_{31}^2: & & E \sim \mathrm{few} \, \mathrm{GeV} \, \mathrm{to} \, 35 \, \mathrm{GeV} \, . \nonumber 
\end{eqnarray}
If the neutrino energy is far out of this range, either the matter effects or the overall event rate
from oscillations will be strongly suppressed. However, there are also possible applications.
Since, for instance, for solar neutrinos $E \ll E_{\mathrm{res}}$, one can use the absence of
the resonance for analytical simplifications, as we will discuss later.

\subsubsection*{Numerical evaluation and conceptual problems}

\begin{figure}[t]
\begin{center}
\includegraphics[width=0.6\textwidth]{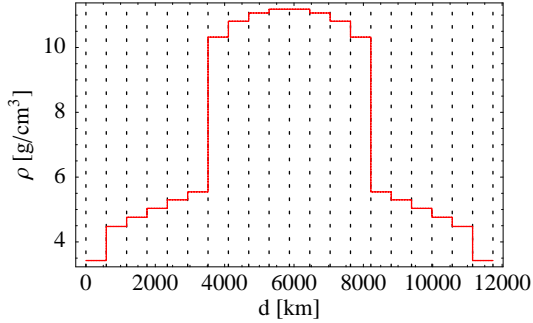}
\end{center}
\caption{\label{fig:profilesliced} Example for a REM- (``Reference Earth Model'') based
matter density profile for a baseline of $12 \, 000 \, \mathrm{km}$
as used for the numerical evaluation of the neutrino propagation
(20~steps). The matter density within each layer is assumed to be constant.
}
\end{figure}

In order to numerically study neutrino oscillation tomography, a commonly
used method is the ``evolution operator method'' (\cf, \eg, \Ref~\cite{Ohlsson:1999um}). 
This method assumes that
the matter density profile be discretized into layers with constant density
(\cf, \figu{profilesliced} for an example). The initial state $| \nu_\alpha \rangle$
is then propagated through the different matter density layers with depths $x_j$ with the
evolution operators
\begin{equation}
\mathcal{V}(x_j,\rho_j) = e^{ - i \mathcal{H}(\rho_j) x_j} \,
\end{equation}
while the
Hamiltonian within each layer $\mathcal{H}$ (\cf, \equ{h}) is assumed to have no explicit time-independence (it is given in constant density $\rho_j$). The transition probability
is then obtained as
\begin{equation}
P_{\alpha \beta} = \left| \langle \nu_\beta | 
\mathcal{V}(x_n,\rho_n) \hdots \mathcal{V}(x_1,\rho_1) | \nu_\alpha \rangle \right|^2 \, .
\label{equ:pevol}
\end{equation}
In practice, \equ{pevol} is evaluated by diagonalizing the Hamiltonian for each density step, \ie, by 
calculating the mass eigenstates in each matter layer. Note that in general
\begin{equation}
[\mathcal{V}(x_i,\rho_i),\mathcal{V}(x_j,\rho_j ) ] \neq 0 \quad \mathrm{for} \quad  \rho_i \neq \rho_j \, ,
\label{equ:commutator}
\end{equation}
which means that the evolution operators of different layers do not necessarily commute.
This already implies that the information from a single baseline must be somehow
sensitive towards the arrangement of the matter density layers. This is very different
from X-ray or neutrino absorption tomography which do not have positional information
from one baseline.

An important conceptual problem in neutrino oscillation tomography is the matter profile
inversion problem~\cite{Ermilova:1986ph,Ermilova:1988pw}. Assume that a matter density profile such as in \figu{profilesliced}
is given. For a specific experiment setup, it is then fairly easy to compute the
corresponding transition probabilities or event rates as function of energy. However,
the reverse problem is theoretically generally unsolved: Assume that the transition probability
is known up to infinite energies, then it would be very useful
to be able to compute the matter profile from that. So far, there have been several
attempts to solve this problem using simplifications, such as
\begin{itemize}
\item
 Simple models using only very few discrete steps (see, \eg, \Refs~\cite{Nicolaidis:1987fe,Nicolaidis:1990jm,Ohlsson:2001fy})
\item
 Linearization in a low density medium (solar, supernova neutrinos)~\cite{Akhmedov:2005yt}
\item 
 Discretization of a more complex profile using non-deterministic
 algorithms to fit a large number of parameters~\cite{Ohlsson:2001ck}.
\end{itemize}
Below, we will discuss some of these approaches in greater detail.

\subsection{Neutrino oscillation tomography with solar and supernova neutrinos}

\begin{figure}[t]
\begin{center}
\includegraphics[width=\textwidth]{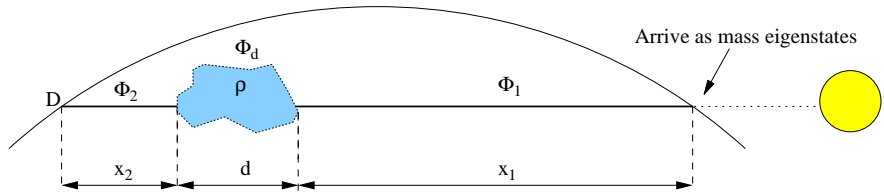}
\end{center}
\caption{\label{fig:sun} Neutrino oscillation tomography
using solar neutrinos for the investigation of a cavity in the Earth's mantle.
}
\end{figure}

Solar and supernova neutrinos are theoretically very interesting for neutrino
tomography because matter effects are introduced off the resonance in Earth
matter, \ie, the neutrino energy $E \ll E_{\mathrm{res}}$ (\cf, \equ{eres}
for $\Delta m^2 = \Delta m_{21}^2$),
or equivalently $\hat{A} \ll 1$. 
This means that one does not expect strong matter effects in Earth matter
as opposed to within the sun. However, this limit is theoretically very useful
to study tomography because it allows for perturbation theory and
other simplified approaches. It is often referred to as neutrino oscillations in a 
``low density medium''~\cite{Ioannisian:2004jk,Ioannisian:2004vv}
 because the density in the Earth is much lower than in the sun.

\subsubsection*{Detecting a cavity}
 
 We show in \figu{sun} a possible setup for neutrino tomography using solar neutrinos
 following \Ref~\cite{Ioannisian:2002yj}. In this setup, the detector is fixed while
 the Earth is rotating, which means that the cavity with density $\rho$  is ``exposed'' 
(in line of sight sun-detector) a time $0 < t_{\mathrm{exp}} < 24 \, \mathrm{hr}$ 
per day. The change in the oscillation probability during this time is, depending on
geometry and density contrast, $\lesssim 0.1 \%$. This leads to a required detector
mass $M \gtrsim 130 \, \mathrm{Mt}/(t_{\mathrm{exp}} [\mathrm{hr}])$, which has a lower limit 
of $5 \, \mathrm{Mt}$ at the poles. Thus, from the statistics
point of view, this approach is very challenging, and backgrounds might be an
important issue. In addition, for such large detectors, the detector surface area might be of the order of the cavity size. There are, however, interesting theoretical results from
such a discussion. Let us define the oscillation phases in the individual steps $x_1$, $d$, and
$x_2$ as~\cite{Ioannisian:2002yj} 
\begin{equation}
\Phi_i \equiv \frac{\Delta m_{21}^2 x_i}{2E} \sqrt{\sin^2 2 \theta_{12} + (\cos 2 \theta -
\hat{A}_i)^2}
\label{equ:solphases}
\end{equation}
with the corresponding matter potentials $\hat{A}_i$ (\cf, \figu{sun}). One can show that if mass eigenstates
arrive at the surface of the Earth (solar and supernova neutrinos), the change in probability
$\Delta P$ (cavity exposed-not exposed) only depends on $\Phi_2$, but not on $\Phi_1$.
In addition, there is a damping of contributions from remote distances $x_2$, which means
that solar neutrinos are less sensitive to the deep interior of the Earth than to
structures close to the detector.

\subsubsection*{Matter density inversion problem}

A further application of the low density limit is to theoretically 
solve the matter profile inversion problem. Following \Ref~\cite{Akhmedov:2005yt}, the Earth matter effect on solar or supernova neutrinos is fully encoded in the quantity (``day-night regeneration effect'')
\begin{equation}
P_{2e}^{night} - P_{2e}^{day} =  \frac{1}{2} \cos^2 \theta_{13} \sin^2 2 \theta_{12} f(\delta)   
\end{equation}
with
\begin{eqnarray}
f(\delta) &  = & \int\limits_0^L dx A(x) \sin \left[ 2 \int\limits_x^L \omega(x') dx' \right]\, ,  
\label{equ:fdelta} \\
\omega(x) & = & \sqrt{(\delta \cos 2 \theta_{12} - A(x)/2)^2 + \delta^2 \sin^2 2 \theta_{12}} \,  , \\
A(x)  & = &  \sqrt{2} G_F n_e(x) \, , \quad \delta \equiv \frac{\Delta m_{21}^2}{4E} \, .
\end{eqnarray}
This implies  that the measured quantity is $f(\delta)$, \ie, a function of energy, 
which needs to be inverted into the matter profile $A(x)$. Especially, the double integral in \equ{fdelta} is quite complicated to invert. However, using the low density limit
$A \ll 2 \delta$ (or equivalently $\hat{A} \ll 1$) as well as $A L \ll 1$ ($L \ll 1 \, 700 \, \mathrm{km}$), one can linearize \equ{fdelta} in order to obtain
\begin{equation}
f(\delta) =  \int\limits_0^L dx A(x) \sin[ 2 \delta (L-x) ] \, .
\end{equation}
This is just the Fourier transform of the matter density profile, \ie, 
\begin{equation}
 A(x) = \frac{4}{\pi} 
\int\limits_0^{\infty} f(\delta) \sin (2 \delta (L-x)) d \delta \, ,
\label{equ:invproblem}
\end{equation}
and the matter density profile inversion problem is solved. One problem is
very obvious from \equ{invproblem}: One needs to know $f(\delta)$ in the
whole interval $0 \le E \le \infty$ which is practically impossible. The
authors of \Ref~\cite{Akhmedov:2005yt} suggest an iteration method to solve this
problem. Additional challenges are statistics and a finite energy resolution,
which is ``washing out'' the edges in the profile. One interesting advantage
of using solar or supernova neutrinos is the sensitivity to asymmetric profiles,
\ie, for mass-flavor oscillations there is no degeneracy between one
profile and the time-inverted version, which otherwise (for flavor-flavor oscillations)
can only be resolved by suppressed three-flavor effects.

\subsubsection*{Supernova neutrinos to spy on the Earth's core}

\begin{figure}[t]
\begin{center}
\includegraphics[width=0.45\textwidth]{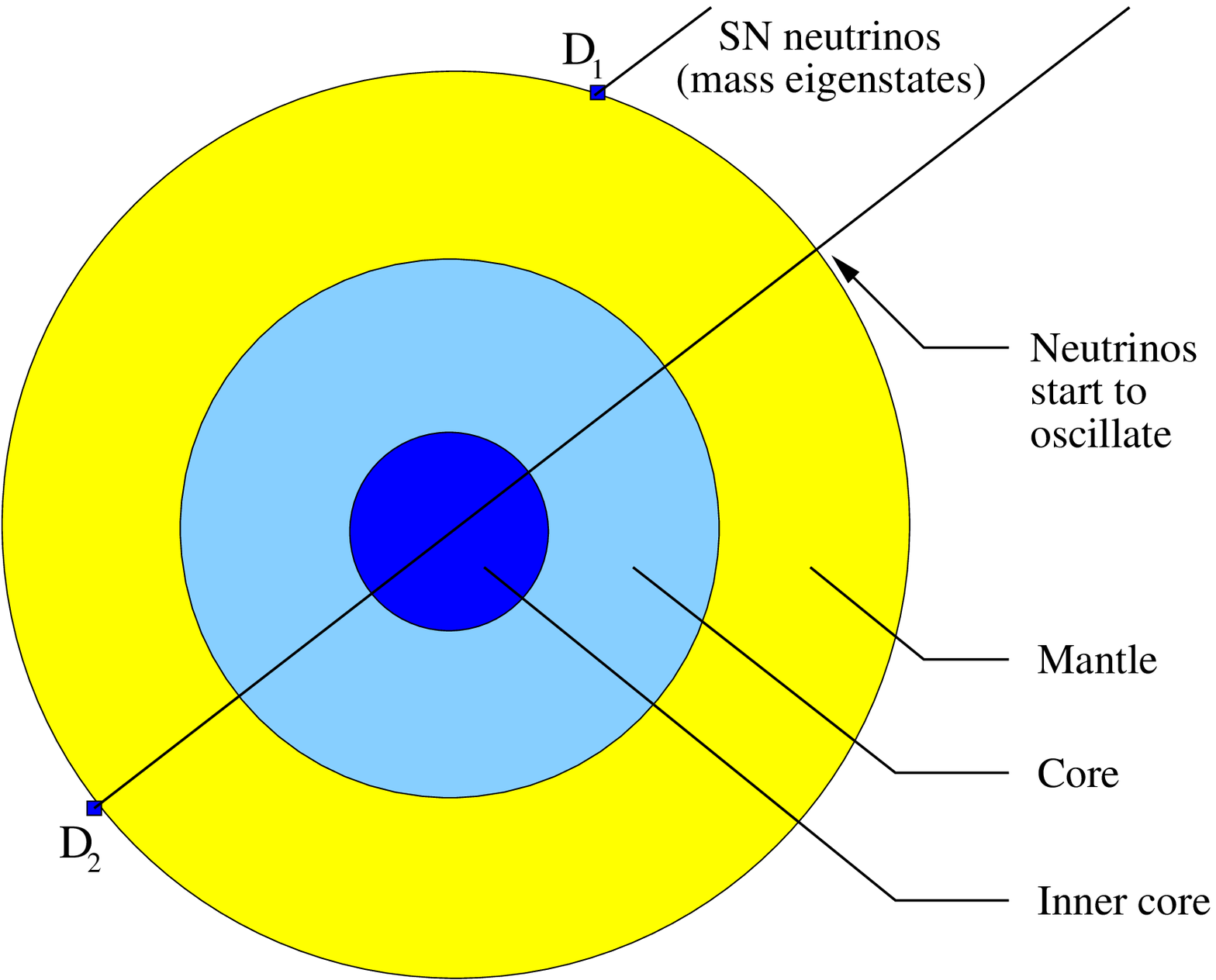}\hspace*{0.1\textwidth}
\includegraphics[width=0.45\textwidth]{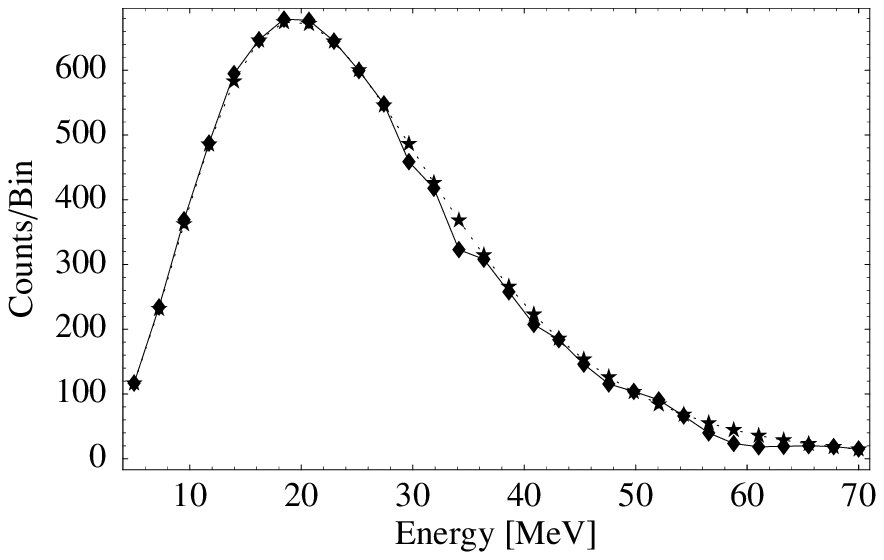}
\end{center}
\caption{\label{fig:supernova} Illustration for the tomography of the Earth's core 
using supernova neutrinos (left) and difference between the event rate spectra
(electron antineutrinos) of $D_1$ and $D_2$ for a Super-Kamiokande-like detectors (right). 
The difference between these spectra corresponds to $\Delta \chi^2 \simeq 35$
for the Earth matter effects, \ie, it is highly significant. 
Figures from \Ref~\cite{Lindner:2002wm}.
}
\end{figure}

Unlike solar neutrinos, which are limited to energies below $20 \, \mathrm{MeV}$,
supernova neutrinos from a possible galactic supernova
explosion have a high-energy tail which is closer to the Earth
matter resonance energy. This effect is illustrated in \figu{supernova} (right)
which compares the energy spectrum between two Super-Kamiokande-like detectors
with and without Earth matter effects. It is obvious from this figure that the
difference between the spectra 
around the peak at $\sim 20 \, \mathrm{MeV}$ is tiny, whereas statistically
significant deviations can be found at larger energies.\footnote{For example,
at around 34~MeV and 60~MeV deviations between the two curves in \figu{supernova} (right)
can be identified. The difference between these spectra corresponds to $\Delta \chi^2 \simeq 35$
for the Earth matter effects, \ie, it is highly significant.} Such a scenario could
happen if supernova neutrinos were detected by two similar-sized detectors, one
on the Earth's surface and with the Earth's core in the line of sight
(\cf, \figu{supernova}, left). Note that the supernova neutrinos are detected 
within a very short time frame $\ll 24 \, \mathrm{hr}$, which means that 
one would actually obtain a ``snapshot'' of the Earth's interior.
As it has been demonstrated in \Ref~\cite{Lindner:2002wm},
for a galactic supernova in the distance $D=10 \, \mathrm{kpc}$ with an energy release of $E=3 \, 10^{53} \, \mathrm{ergs}$, two megaton-size water Cherenkov detectors could measure
the density of the Earth's core at the per cent level with a number of challenges:
First, the Earth's mantle density is assumed to be known at the 2\% level. Second, the
solar oscillation parameters have to be known at the $0.2\%$ level. Third, too similar
supernova fluxes for the different flavors (similar temperatures) and deviations from
energy equipartition are unfavorable. And fourth, one has to have some knowledge on
the flavor composition of the flux, possibly from detection of different flavors.

\subsection{Neutrino oscillation tomography with neutrino beams}

We now discuss neutrino oscillation tomography with the ``man-made'' neutrino beams.
Neutrino beams are planned or future neutrino sources using accelerators, where the neutrino beam is 
produced by pion/kaon decays (superbeams, see, \eg , \Refs~\cite{Itow:2001ee,Ayres:2004js}), by muon decays (neutrino factory, see, \eg , \Refs~\cite{Geer:1998iz,Apollonio:2002en,Albright:2004iw}), or by the decay of unstable nuclei ($\beta$-Beam, see, \eg , \Refs~\cite{Zucchelli:2002sa,Bouchez:2003fy,Burguet-Castell:2003vv,Huber:2005jk}).
Neutrino beams have, compared to ``natural'' neutrino sources, the advantage that either 
flux and flavor composition are well-known, or a near detector can be used to improve the
knowledge on these quantities as well as on the interaction cross sections. There is, however,
one major obstacle common to all of these experiments: Matter effects especially enter
in the $\nu_e \leftrightarrow \nu_\mu$ flavor transition which is suppressed by the small mixing angle $\stheta$. Up to now, this mixing angle is unknown and only an upper bound exists~\cite{Apollonio:1999ae}. Experiments within the coming ten years will reveal if $\stheta$ is suitably large for the applications discussed here (for a summary, see, \eg, \Ref~\cite{Huber:2004ug}). Therefore, the experiment performance has always to be evaluated as function of $\stheta$. In this section, we split the discussion into conceptual areas linked to tomography with neutrino beams.

\subsubsection*{Positional information for a single baseline}

Interesting questions are discussed in \Ref~\cite{Ohlsson:2001fy}: Assume
we have a beam crossing a cavity with a specific density
contrast compared to the surrounding matter. Then one wants to know
\begin{itemize}
\item 
 How large has the cavity to be to be detected?
\item 
 Can the position of the cavity be measured and if so, how precisely?
\end{itemize}
In \Ref~\cite{Ohlsson:2001fy} a $500 \, \mathrm{MeV}$ superbeam is assumed with very luminous $200 \, 000$ events in total. The density in the cavity is assumed to be $1 \, \mathrm{g/cm^3}$ (water), the baseline $L=1 \, 000 \, \mathrm{km}$, and $\stheta=0.03$, where a smaller number of events can be compensated by a larger $\stheta$. It turns out that the cavity has to be longer than about $100 \, \mathrm{km}$ to be found and its size can be measured to about $\pm 50 \, \mathrm{km}$. The most
important result is that the position of the cavity can be reconstructed $\pm 100 \, \mathrm{km}$ from a single baseline, which is very different from X-ray or absorption tomography. However, there is a degeneracy in the position between $x$ and $L-x$ which can be only resolved by suppressed three-flavor effects.  This example demonstrates
already one of the basic principles of neutrino oscillation tomography: Positional information is available already from a single baseline.

\subsubsection*{Resolution of structures and edges}

\begin{figure}[t]
\begin{center}
\includegraphics[width=\textwidth]{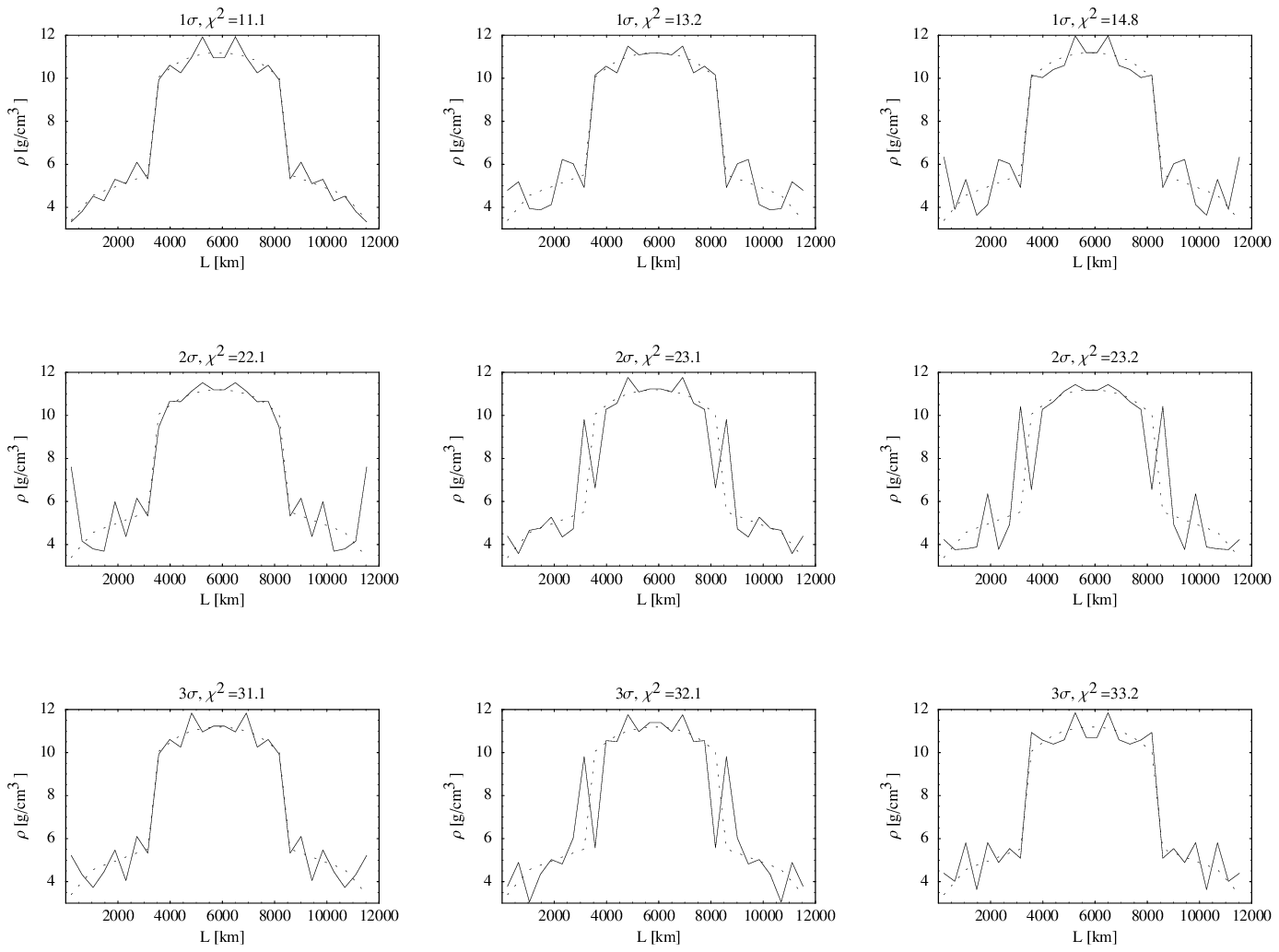}
\end{center}
\caption{\label{fig:samplespace} Examples for reconstructed
(symmetric) REM profiles from an extremely luminous neutrino factory close
to the $1\sigma$ (upper row), $2\sigma$ (middle row), and 
$3 \sigma$ (lower row) contours for $L=11 \, 736 \, \mathrm{km}$
(14 d.o.f.).
Figure from \Ref~\cite{Ohlsson:2001ck}.
}
\end{figure}

One can learn about the resolution of structures and edges from the numerical
solution of the matter density inversion problem. In \Ref~\cite{Ohlsson:2001ck} 
a (symmetrized) REM profile is reconstructed from a single baseline crossing the outer core 
with 14~degrees of freedom using a genetic algorithm. Naturally, there are many 
degenerate profiles close to the $1\sigma$, $2 \sigma$, and $3 \sigma$ contours,
and one cannot show a contour in 14-dimensional parameter space. Therefore, we
show in \figu{samplespace} several ``typical'' representatives close to the $1\sigma$, $2 \sigma$, and $3 \sigma$ contours for a neutrino factory, where the total number of oscillated events is for $\stheta=0.1$ only about a factor of four above currently discussed luminosities (see, \eg, \Ref~\cite{Huber:2002mx}). From \figu{samplespace}, one can easily read off that
such an experiment could, in principle, reconstruct the mantle-core-mantle structure of the
Earth. However, structures smaller than several hundred kilometers cannot be resolved.
In addition, the mantle-core boundary cannot be resolved at a sufficiently high confidence level
from a single baseline. Analytically, it has been demonstrated in \Ref~\cite{Ohlsson:2001ck} that
structures much smaller than the oscillation length in matter cannot be resolved -- as one would naturally expect similar to other quantum mechanical phenomena. In conclusion, neutrino oscillations in matter are very sensitive towards average densities and the arrangement of structure on the length scale of the oscillation length. However, neither can edges nor small structures be precisely resolved.

\subsubsection*{Density measurement}

\begin{figure}[t]
\begin{center}
\includegraphics[width=0.6\textwidth]{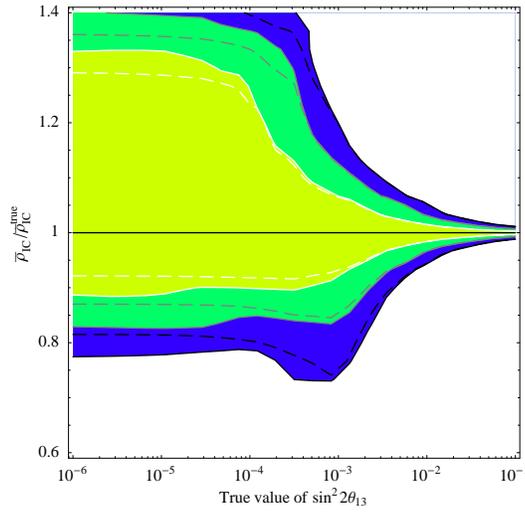}
\end{center}
\caption{\label{fig:th13dep}  The measurement of $\bar{\rho}_{IC}$ (inner core
density) as function of the true value of $\stheta$ at the $1\sigma$, $2\sigma$, and $3 \sigma$ confidence levels (from light to dark shaded regions).
For the baselines, $L=2 \cdot R_E$ combined with a shorter baseline $L=3 \, 000 \, \mathrm{km}$ 
to reduce correlations is used. The
dashed curves correspond to fixing the oscillation parameters, \ie, to not taking into
account correlations and degeneracies.
Figure from \Ref~\cite{Winter:2005we}.
}
\end{figure}

Since we know that neutrino oscillations measure more or less the baseline-averaged densities $\bar{\rho}^L_i = 1/L \int_0^{L} \tilde{\rho}(l) dl$ over long distances plus some suppressed interference effects, we can use this to discuss possible applications. For example, let us 
assume that we want to
perform a simple one-parameter measurement of the Earth's inner core density. Because the Earth's mass is fixed, we need to correct the average mantle or outer core density for any shift of the inner core density. Note, however, that it is the volume-averaged density to be corrected, which means that large shifts in the Earth's inner core density cause only very small density corrections in the mantle. This example illustrates already one potential strength of neutrino oscillation tomography: Since neutrinos from a ``vertical'' baseline travel similar distances in mantle, core, and inner core, there should be no {\em a priori} disadvantage for the innermost parts of the Earth.
In \Ref~\cite{Winter:2005we} a neutrino factory setup from \Ref~\cite{Huber:2002mx} with currently anticipated luminosities was chosen to test this hypothesis for realistic statistics. In order to measure the oscillation parameters, the experiment with $L=2 \, R_E$ was combined with a $L=3 \, 000 \, \mathrm{km}$.
The precision of the measurement can be found in \figu{th13dep} as function of $\stheta$. One case easily read off that a per cent level measurement is realistic for $\stheta \gtrsim 0.01$. Most importantly, the application survives the unknown oscillation parameters and the performance is already close to the optimum (dashed curves). For smaller values of $0.001 \lesssim \stheta \lesssim 0.01$, the correlations would be much worse without the $L=3 \, 000 \, \mathrm{km}$ baseline. For large values of $\stheta \gtrsim 0.01$, the vertical baseline alone is hardly affected by correlations with the oscillation parameters: As illustrated in \Ref~\cite{Winter:2005ud},
CP effects are suppressed for very long baselines. 
\begin{figure}[t]
\begin{center}
\includegraphics[width=\textwidth]{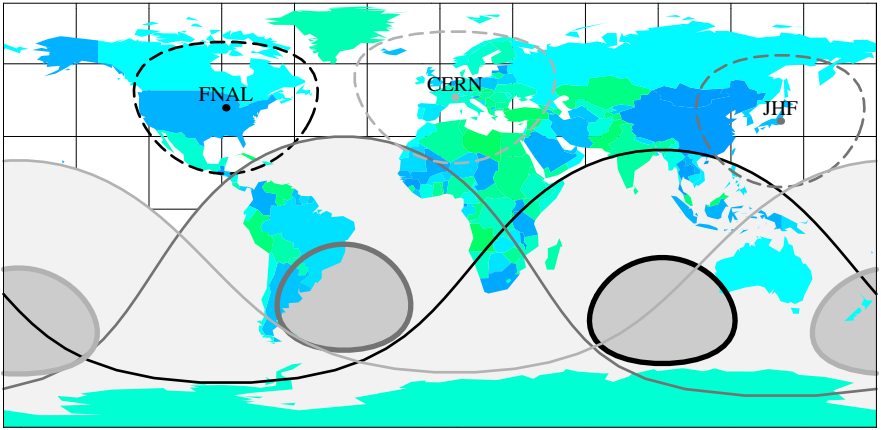}
\end{center}
\caption{\label{fig:geography} Positions of three of the major potential neutrino
factory laboratories, (typical) $L=3 \, 000 \, \mathrm{km}$ detector sites (dashed curves),
as well as
potential detector sites with outer core crossing baselines (below thin solid curves), and inner core crossing baselines (within thick solid curves). The colors of the curves
represent the different laboratories.
Figure from \Ref~\cite{Winter:2005we}.
}
\end{figure}
Since there is only a number of potential high-energy laboratories around the world which could
host a neutrino factory, we show in \figu{geography} some examples and the corresponding outer and
inner core crossing baselines. Obviously, there are potential detector locations for some of the
laboratories, which are, however, not exactly on the $L=2 R_E$-axis. Relaxing this baseline constraint
somewhat, one can show that one can find detector locations for a small drop in precision~\cite{Winter:2005we}. In summary, this application illustrates that a density measurement
could be performed with a) reasonable statistics, b) including the correlations with the oscillation parameters, and c) reasonably small values of $\stheta$. In the future, it has to be clarified 
how large the additional effort for such a facility (the vertical storage ring) would be.
Note, however, that there are plenty of other applications of a ``very long'' neutrino factory baseline, such as the ``magic baseline'' to resolve degeneracies~\cite{Huber:2003ak} ($L \sim 7 \, 500 \, \mathrm{km}$), the test of the MSW effect for $\stheta=0$~\cite{Winter:2004mt} ($L \gtrsim 5 \, 500 \, \mathrm{km}$), the mass hierarchy measurement for $\stheta=0$~\cite{deGouvea:2005hk,deGouvea:2005mi} ($L \sim 6 \, 000 \, \mathrm{km}$),  and the test of the ``parametric resonance''~\cite{Akhmedov:1998ui,Petcov:1998su} ($L \gg 10 \, 665 \, \mathrm{km}$).

In addition to the described neutrino sources, note that tomography comparing the
neutrino and antineutrino disappearance information from atmospheric neutrinos might, in principle, be possible as well~\cite{GeiserKahle}.

\section{Other geophysical aspects of neutrino oscillations}

\begin{figure}[t]
\begin{center}
\includegraphics[width=0.6\textwidth]{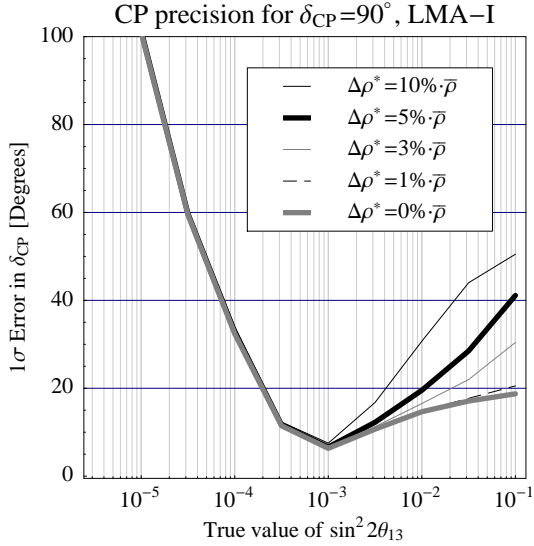}
\end{center}
\caption{\label{fig:cpprecmatter} The precision of the measurement
  of $\deltacp$  for a neutrino factory and the simulated value
  $\deltacp=90^\circ$ as a function of the true value of $\stheta$ at
  the $1\sigma$ confidence level. The different curves correspond to
  different allowed matter density uncertainties $\Delta \rho^*$ as
  described in the plot legend, especially the thick curves correspond to
  no matter density uncertainty (light thick curve) and the often used
  uncertainty $\Delta \rho^* = 5 \% \cdot \bar{\rho}$ (dark thick
  curve).
Figure from \Ref~\cite{Ohlsson:2003ip}.
}
\end{figure}

It is well known that matter
density uncertainties spoil the extraction of the oscillation parameters from the
measurements (see  \Refs~\cite{Jacobsson:2001zk,Geller:2001ix,Shan:2001br,Fogli:2001tm,Ota:2002fu,Shan:2003vh,Kozlovskaya:2003kk,Ohlsson:2003ip} and references therein).
In particular for baselines sensitive to $\deltacp$, such as $L \sim 3 \, 000 \, \mathrm{km}$ at a 
neutrino factory, the additional correlation with the matter density affects the
precision measurements of $\stheta$ and $\deltacp$, and the CP violation sensitivity. This effect is illustrated in \figu{cpprecmatter} for the precision of $\deltacp$ and different matter density uncertainties $\Delta \rho^*$. Especially for large $\stheta$, any uncertainty larger than about $1\%$ affects the precision severely. Note that the baseline used for \figu{cpprecmatter} 
is $L=3 \, 000 \, \mathrm{km}$, which means that the neutrinos travel in an average depth of $\sim 120 \, \mathrm{km}$ up to a maximum depth of $\sim 180 \, \mathrm{km}$. In these depths,
the uncertainty among geophysics models is currently at the level of 5\%~\cite{Geller:2001ix}.
Since the matter density uncertainties may affect the competitiveness of a neutrino factory
with a superbeam (operated at shorter baselines) for large values of $\stheta$, improved knowledge for specifically chosen baselines would be very helpful.

\section{Summary and conclusions}

In summary, neutrino tomography might be a very complementary approach to geophysical methods.
For example, neutrinos travel on straight lines with almost no uncertainty in their path. Furthermore, neutrino tomography is either sensitive to the nucleon density (absorption tomography) or electron density (oscillation tomography). In comparison, the paths of seismic waves are curved, and there is some uncertainty in them. In addition, the matter density has to be reconstructed from the propagation velocity profile by the equation of state. This means that neutrino tomography might be a more ``direct'' handle on the matter density and could be very useful to investigate specifically localized regions, such as in the lower mantle. Moreover, there is no principle
reason to prevent neutrinos from penetrating the Earth's core, whereas seismic waves are partially reflected at the mantle-core and outer-inner core boundaries. Note that though the most precise information on deviations from the REM (``Reference Earth Model'') in the Earth's mantle comes from seismic waves, there are other geophysical methods which might be more directly sensitive towards the matter density, such as normal modes, mass, and rotational inertia of the Earth. Nevertheless, none of those could provide a measurement along a very specific path.

The main challenges for neutrino tomography might be the existence of high-energy neutrino sources for absorption tomography, and the statistics for oscillation tomography.
For example, neutrino oscillation tomography could, in principle, reconstruct the matter
density profile along a single baseline due to interference effects among different matter
density layers. Note, however, that neutrino oscillations
are to first order sensitive towards densities averaged over the scale of the oscillation length,
which means that such sophisticated applications require extremely large
luminosities (detector mass $\times$ source power $\times$ running time) and might be very challenging. On the other hand, 
very simple questions, such as a one-parameter measurement of the average density along the
path or the discrimination between two very specific degenerate geophysical models might be
feasible within the next decades. For example, the achievable precision for the inner
core density of the Earth with a neutrino factory experiment might be
quite comparable ($\pm 0.23 \, \mathrm{g/cm^3}$ for $\stheta=0.01$ and
 $\pm 0.06 \, \mathrm{g/cm^3}$ for $\stheta=0.1$~\cite{Winter:2005we}) 
 to current precisions given for the density jump at the inner-core 
boundary from geophysics (\eg, $\pm 0.18 \, \mathrm{g/cm^3}$ in \Ref~\cite{Masters}).
We therefore conclude that it will be important that the right and simple questions be asked by discussions between neutrino physicists and geophysicists.

\begin{acknowledgements}
I would like to thank Evgeny Akhmedov, Tommy Ohlsson, and Kris Sigurdson for useful
comments and discussions. In addition, I would like to thank the organizers of
the workshop ``Neutrino Sciences 2005: Neutrino geophysics'' for an excellent workshop
with vivid discussions.
\end{acknowledgements}

\end{document}